\documentclass[preprint,5p,times,twocolumn]{elsarticle}
\usepackage{amssymb}
\usepackage{amsmath}
\usepackage{epsfig}
\journal{ Chin. Phys. C}
\begin{document}
\begin{frontmatter}
\title{The unified Skyrmion profiles and Static Properties of Nucleons}
\author{Duo-Jie Jia$^{\ast}$}
\ead{jiadj@nwnu.edu.cn}
\author{Xiao-Wei Wang}
\cortext[ast]{Corresponding author}
\address{Institute of Theoretical Physics, College of Physics
and Electronic \\Engineering, Northwest Normal University,\textit{\
Lanzhou 730070, P.R. China}}

\begin{abstract}
An unified approximated solution for symmetric Skyrmions was
proposed for the SU(2) Skyrme model for baryon numbers up to 8,
which take the hybrid form of a kink-like solution and that given by
the instanton method. The Skyrmion profiles are examined by
computing lowest soliton energy as well as the static properties of
nucleons within the framework of collective quantization, with a
good agreement with the exact numeric results. The comparisons with
the previous computations as well as the experimental data are also
given.
\end{abstract}

\end{frontmatter}
\section{Introduction}

Due to its substantial connections with the baryon phenomenology,
the Skyrme model\cite{Skyrme}, which appears as the effective field
theory of QCD at
low-energy limit, has been visited extensively since 1980's (see, \cite%
{Zahed}, for a review), of which the soliton solution, known as
Skyrmion, has been often used to explore the baryon and barynic
system spectrum and their static properties, particularly, in the
case of the small baryon numbers ($B$)
\cite{Sutcliffe,Battye97,AtiyahManton,Atiyah89,Battye}. Owing to the
high nonlinearity, the most of Skyrmion studies was confined to the
numerical approach \cite{ANW,AN,Sutcliffe,Battye97,Battye}. It is
still worthwhile, however, to seek the approximated analytic
soluitions of Skyrmions considering its extensive applications in
baryon phenomenology and nucler physics. The some of efforts have
been made in finding the
approximated solution to the massless Skyrme model \cite%
{JMP,Atiyah89,Ponciano,Yamashita,JWcpl}, but few has been done to
the massive Skyrme\newline model, except for the exploration by
Manton and Atiyah \cite{SKatiyah2005}.

In this Letter, we revisit the static solutions of Symmetric
Skyrmions in the $SU(2)$ massive Skyrme model, which includes the
chiral symmetry breaking term (namely, pion mass term). An
analtyical Skyrmion profile is proposed for the massive Skyrme mode
with the baryon number up to $8$ by writing it as the hybrid form of
a kink-like solution and the analytic solution obtained by the
instanton method \cite{Atiyah89}. We show that this profile is in
well consistent with the exact numeric solution obtained via the
standard relaxation method. To determine the parameters involved in
the profile, the downhill simplex algorithm is applied in lowering
the energy as best as possible. The static properties of nucleon as
well as delta are then
computed using the $B=1$ solution in the case of the reduced pion mass $%
m=0.526$ within the framework of the semiclassical quantization of
the Skyrme model, and compared to the previous computations and
experimental data.

\section{The analytical profiles for Skyrmions}

The $SU(2)$ Skyrme\ model \cite{Skyrme} with pion mass term is given
by
\begin{equation}
\begin{split}
L=\int d^{3}x\left\{\frac{f{}_{\pi }^{2}}{4}tr\left(\partial _{\mu
}U\partial ^{\mu }U^{\dag }\right)+\frac{1}{32e^{2}}tr\left[\partial
_{\mu }UU^{\dag },\partial _{\nu
}UU^{\dag }\right]^{2} \right. \\
\left.+\frac{m_{\pi }^{2}}{2}f_{\pi }^{2}tr(U-1)\right\}%
\end{split}
\label{SK}
\end{equation}%
with $U(x,t)\in SU(2)$ the nonlinear realization of the chiral field
describing the $\sigma $ field and $\pi $ mesons under the constrain $%
U^{\dag }U=1$, $2f_{\pi }$ is the pion decay constant, and $e$ is
the dimensionless constant characterizing nonlinear coupling. The
last (mass) term in (\ref{SK}) corresponds to the chiral symmetry
breaking and ignored in the chirally symmetric Skyrme model. With
the weak-$\pi $ expansion
\begin{equation*}
U=(\sigma +i\vec{\tau}\cdot \vec{\pi})/f_{\pi }\approx 1+\frac{i}{f_{\pi }}%
\vec{\tau}\cdot \vec{\pi},
\end{equation*}%
($\vec{\tau}$ are the three Pauli matrices) the model (\ref{SK})
becomes the dynamics of pian
\begin{equation*}
\mathcal{L}^{\pi }=\frac{1}{2}(\partial _{\mu
}\vec{\pi})^{2}-\frac{m_{\pi }^{2}}{2}\vec{\pi}^{2}+\mathcal{O}(\pi
^{4}),
\end{equation*}%
in which the pion-mass term arises from the mass (third) term in
(\ref{SK}). The Hamiltonian with respect to the Lagrangian
(\ref{SK}) is
\begin{equation}
\begin{split}
\mathcal{H}=\frac{f_{\pi }}{2e}\left\{\frac{1}{2}tr\left(\partial
^{i}U\partial ^{i}U^{\dag }\right)-\frac{1}{16}tr\left[\partial
^{i}UU^{\dag },\partial ^{j}UU^{\dag
}\right]^{2}\right. \\
\left.+m^{2}tr\left(1-U\right)\right\},%
\end{split}
\label{SK1}
\end{equation}%
with $m=m_{\pi }/(ef_{\pi })$ the reduced pian mass (in the unit of
$ef_{\pi
}$), and $\partial ^{i}$ are the partial derivatives in the length unit of $%
1/ef_{\pi }$.

We use the parameterization of the rational-map ansatz described in \cite%
{RationM98},
\begin{equation}
U(r,z)=\exp \left[iF(r)\hat{n}_{R(z)}\cdot \vec{\tau}\right]
\label{RM0}
\end{equation}%
where $\hat{n}_{R(z)}$ represents a map from the Riemann spherical
coordinate $z=e^{i\varphi }\tan (\theta /2)$, which is related to
the angular coordinate $(\theta ,\varphi )$ via the stereographic
projection, to the unit iso-triplet through the next stereographic
projection in the internal space,
\begin{equation}
\hat{n}_{R(z)}=\frac{1}{1+|R|^{2}}\left( 2\textrm{Re} R,2\textrm{Im}%
R,1-|R|^{2}\right) .  \label{RM}
\end{equation}%
Here, the map $R(z)$ from a complex plane to $S^{2}=\mathbf{C}\cup
\{\infty \}$ in the internal space has also to be a holomorphic map
of degree $B$.
This implies that $R(z)$ can be written as a ratio of two polynomials $%
R(z)=p(z)/q(z)$, where $p$ and $q$ are polynomials in $z$ such that $%
max\{deg(p),deg(q)\}=B$, and $p$ and $q$ have no common factors. In (\ref%
{RM0}), $F(r)$ is a real profile function of the radial coordinate
$r$,
which has to satisfy the boundary condition $F(0)=\pi ,F(\infty )=0$. When (%
\ref{RM0}) and (\ref{RM}) applied, $U$ becomes
\begin{equation}
U(r,z)=\frac{1}{1+|R|^{2}}\left[
\begin{array}{cc}
e^{iF}+|R|^{2}e^{-iF} & 2iR^{\ast }\sin (F) \\
2iR\sin (F) & e^{-iF}+|R|^{2}e^{iF}%
\end{array}%
\right]  \label{Rmap}
\end{equation}%
With (\ref{Rmap}) and (\ref{SK1}), one has for the static energy for (\ref%
{SK})
\begin{equation}
M^{SK}=6\pi^{2} \frac{f_{\pi }}{e}E
\end{equation}%
\begin{equation}
\begin{split}
E=\frac{1}{3\pi } \int_{0}^{\infty }dxx^{2}\left[F_{x}^{2}+%
\frac{2}{x^{2}}B\sin ^{2}(F)(1+F_{x}^{2})\right.\\
\left.+\mathcal{I}\frac{\sin ^{4}(F)}{%
x^{4}}+2m^{2}(1-\cos F)\right]
\end{split}
\label{Eden}
\end{equation}%
where
\begin{equation}
\mathcal{I}=\frac{1}{4\pi }\int \left(
\frac{1+|z|^{2}}{1+|R|^{2}}\left\vert \frac{dR}{dz}\right\vert
\right) ^{4}\frac{2idzdz^{\ast }}{(1+|z|^{2})^{2}} \label{I}
\end{equation}%
Here, the dimensionless variable $x=ef_{\pi }r$ and $F_{x}\equiv
dF(x)/dx$ are used. The equation of motion of (\ref{Eden}) is
\begin{equation}
\begin{split}
\left( 1+2B\frac{\sin ^{2}F}{x^{2}}\right) F_{xx}+\frac{2}{x}F_{x}
+\frac{\sin (2F)}{x^{2}}\\\times\left( F_{x}^{2}-1-\mathcal{I}\frac{\sin ^{2}F}{x^{2}%
}\right)  =m^{2}\sin F,%
\label{EOM}
\end{split}
\end{equation}%

The numerical solutions to (\ref{EOM}) and to the equation of motion
of the original functional (\ref{SK1}) via the full field $U$
calculation are both presented by Battye, Sutcliffe and others
\cite{Sutcliffe}. It is shown there that an excellent agreement was
achieved between the solutions given by rational map approximation
and full field calculation for the baryon number $B$ up to $12$.
Here, we re-examine only the rational map solutions
to the Skyrmions with $B=1\sim 8$ for the case of the reduced pion mass $%
m=0.526$, which is of the physically interest \cite{AN}, by solving (\ref%
{EOM}) analytically and numerically.

For given $B$, the minimization of the energy (\ref{Eden}) should be
carried
out in two steps: to minimize $\mathcal{I}$ (\ref{I}) as a functional of $%
R(z)$ at first and then to minimize energy (\ref{Eden}) for the
optimized value of $\mathcal{I}$. The first step is done in
\cite{RP} and we simply list optimized values of $\mathcal{I}$ in
\textrm{Table I} as a input for our calculation.

The Skyrmion profile ansatz we propose is
\begin{equation}
\begin{split}
F(x)=4w\arctan \left[\exp (-cx)\right]+
\pi (1-w)\\\times\left[ 1-\left( \frac{\sinh ^{2}(dmx)}{%
a^{2}+\sinh ^{2}(dmx)}\right) ^{1/2}\right]
\end{split}
\label{Fw}
\end{equation}%
where $m\neq 0$, and $(a,c,d,w)$ are the parameters to be determined
variationally. We note that the the second term in (\ref{Fw})
appears as the
hyperbolic Skyrmion form obtained using instanton holonomy\cite%
{Manton1990,SKatiyah2005} where $a$ is related to the instanton scale $%
\lambda $ through
\begin{equation}
a=\frac{2\lambda }{(1+\lambda ^{2})}.  \label{a}
\end{equation}%

Given $F(x)$ specified by (\ref{Fw}), we are able to minimize
(\ref{Eden}) with respect to the variational parameters ($a,c,d,w$)
for $B=1,\cdots ,8$ using the downhill simplex method (the
Neilder-Mead algorithm). The optimized parameters are listed in
\textrm{Table I}, including the soliton energy per $B$ calculated
accordingly and the corresponding numeric results obtained via the
relaxation procedure.

The profile solution (\ref{Fw}), with the parameters in
\textrm{Table I}, is plotted in \textrm{Fig.1} and compared to the
numerical solution to the equation (\ref{EOM}). To the later, we
employ the relaxation algorithm with the different initial setups of
$F(x)$, ranging from the Gaussian distribution, to the kink
configuration, and find that the same result is achieved. It can be
seen from \textrm{Table I} and \textrm{Fig.1} that a quite well
agreement is achieved between the numerical solutions and that given
by (\ref{Fw}).

We also present the fit of the parameters depending upon $B$, which
are given by
\begin{equation}
\begin{array}{l}
\lambda =0.4639+\allowbreak 1.488\exp \left( 15.038-15.361B\right) \\
-0.5466\exp \left( 1.7442-1.7816B\right) , \\
c= 0.7196+2.\,\allowbreak 287\,1\frac{B-0.948\,45}{B-1\allowbreak 05.5\,9}%
\exp \left\{ -\left[ \ln \left( \frac{B}{0.2236}\right) \right]
^{2}/6.774\right\}, \\
d=0.7981B^{\left[ 0.114\,12\left( \frac{13.\,\allowbreak 056-B}{B-0.92643}%
\right) ^{0.7863}\right] }, \\
w=251.4-\frac{248.8}{1+\exp \left\{ 21.\allowbreak 394\frac{B-0.467\,09}{%
B+0.06651\,9}-20.404\right\} },\label{param}%
\end{array}%
\end{equation}
and shown in \textrm{Fig.3}. $\allowbreak $

For the dependence of the charge one Skyrmion profile on the reduced
pion mass $m$, we examine (\ref{Fw}) for the different $m=1,2,3$,
respectively. In \textrm{Table II}, the list is given for the
parameters in (\ref{Fw}), which\ are calculated with the same
procedure as before, and for the corresponding soliton energies
obtaind by (\ref{Fw})(marked by Ana) and the energies obtained
numerically (marked by Num). The chiral angle profiles obtained
analytically and numerically are both plotted in \textrm{Fig. 2} for
$m=0,1,2,3$, where the Skyrmion profile for the massless model is
otained in \cite{JWcpl} and used simply here for $m=0$ case. The
results here show that the function (\ref{Fw}) for $B\geq 2$ can be
taken to be the approximated analytical solution to the Skyrme
model, which is presented in an analytical form for any low $B$.
Since (\ref{Fw}) are not the most general expression of $F(x)$, the
variation of the parameters does not necessarily lead to a real
analytical solution. However, considering that exact analytical
solution is difficult (almost impossible) to be found in most cases
of the highly nonlinear problems like the Skyrme model, the
solutions presented here can be an alternative to the exact
analytical solution in so far as we are going to search the
approximated solution which approaches the real solution as better
as possible.

\section{The static properties of nucleons}

It is known that the static properties of nucleons can be extracted
by semi-classically quantizing the spinning modes of Skyrme
Lagrangian using the collective variables\cite{ANW,AN}. Following
Adkin et al. \cite{AN}, we use the charge one solution of (\ref{Fw})
to compute the static properties of nucleons and
nucleon-isobar($\Delta $) within the framework of the bosonic
quantization of Skyrme model.

Choosing a $SU(2)$-variable $A(t)$ as the collective variables, and
substituting $U=A(t)U_{0}(x)A(t)^{\dagger }$ into (\ref{SK}), one
can show that the Hamiltonian for the bosonically quantizatized
model has an eigenvalue $\langle H\rangle
=M^{SK}+J(J+1)/(2I_{0}\Lambda )$ with $M^{SK}$
the soliton energy for the static hedgehog Skyrmion($\hat{n}_{R(z)}=\hat{r}$%
) for which $\mathcal{I}=1$, $I_{0}=\pi /(3e^{3}f_{\pi })$, and
\begin{equation}
\Lambda =8\int_{0}^{\infty }x^{2}dx\sin ^{2}F\left[ 1+F_{x}^{2}+\sin
^{2}F/x^{2}\right] .  \label{lam}
\end{equation}%
Hence, the masses of the nucleon and $\Delta $-isobar are given by (with $%
J=1/2$ and $3/2$)
\begin{equation}
\begin{split}
M_{N}& =M^{SK}+\frac{3}{8I_{0}\Lambda }, \\
M_{\Delta }& =M^{SK}+\frac{15}{8I_{0}\Lambda }. \label{ndm}
\end{split}%
\end{equation}%
As done in \cite{AN}, we choose to adjust model parameters ($f_{\pi
},e$) to fit the hadron masses, namely, the $N$, $\Delta $ , and
$\pi $ masses
through (\ref{ndm}), where the other quantities involved are given by (\ref%
{Eden}) and (\ref{lam}). Our reports for the computation of the
static properties of baryons are presented in \textrm{Table III},
including the experimental data as well as the results in
Ref.\cite{ANW,AN}. The computations parallel the computation
reported in \cite{AN}. We also list the most relevant formulas for
our computations, for instance, the isoscalor root mean
square(r.m.s) radius and isoscalor magnetic r.m.s radius

\begin{equation}
\begin{split}
ef_{\pi }\langle r^{2}\rangle _{I=0}^{1/2}=& \left\{ -\frac{2}{\pi }%
\int_{0}^{\infty }x^{2}\sin ^{2}FF_{x}\right\} ^{1/2} \\
ef_{\pi }\langle r^{2}\rangle _{M,I=0}^{1/2}=& \left\{ \frac{%
\int_{0}^{\infty }x^{4}\sin ^{2}FF_{x}dx}{\int_{0}^{\infty
}x^{2}\sin ^{2}FF_{x}dx}\right\} ^{1/2}
\end{split}
\notag
\end{equation}%
and the magnetic moments for proton and neutron
\begin{equation}
\mu _{p,n}=\mu _{p,n}^{I=0}+\mu _{p,n}^{I=1}=\frac{\langle
r^{2}\rangle _{I=0}}{9}M_{N}(M_{\Delta }-M_{N})\pm
\frac{M_{N}}{2(M_{\Delta }-M_{N})}, \label{mu}
\end{equation}%
where plus and minus correspond to proton and neutron, respectively.
Since the chiral symmetry is broken explicitly, the isovector
electric and
magnetic charge radii are finite and can be calculated by%
\begin{equation}
\begin{split}
e^{2}f_{\pi }^{2}\langle r^{2}\rangle _{I=1}=& \int_{0}^{\infty
}x^{2}\rho
_{I=1}(x)dx \\
e^{2}f_{\pi }^{2}\langle r^{2}\rangle _{M,I=1}=& \int_{0}^{\infty
}x^{2}\rho _{M,I=1}(x)dx
\end{split}
\label{Isradii}
\end{equation}%
where $\rho _{I=1}(x)$ is the normalized electric iso-vector charge
distributions and can be derived from the zero component of the
vector current. It can be shown that these two densities are
identically same and can be given by
\begin{equation*}
\rho _{I=1}(x)=\rho _{M,I=1}(x)=\frac{x^{2}\sin ^{2}(F)\left\{
1+F_{x}^{2}+\sin ^{2}(F)/x^{2}\right\} }{\int_{0}^{\infty }x^{2}\sin
^{2}(F)\left\{ 1+F_{x}^{2}+\sin ^{2}(F)/x^{2}\right\} dx}.
\end{equation*}%
Therefore, the r.m.s radii of the isovector electric and magnetic
charge are happen to be same.

For the nuclear interaction, the pian-nucleon sigma term can be
given by the matrix element in the nucleon states $N(p)$
\cite{Pagels75}, which in our case becomes (see \cite{AN})
\begin{equation*}
\sigma =\frac{1}{2}f_{\pi }^{2}m_{\pi }^{2}\int_{0}^{\infty }dV[2-trU_{0}] \\
=\frac{4\pi m_{\pi }^{2}}{e^{3}f_{\pi }}\int_{0}^{\infty
}x^{2}dx\left( 1-\cos (F)\right) ,
\end{equation*}%
where $U_{0}$ is the the soliton solution of the chiral symmetric
theory(namely, the massless theory). The result is $\sigma
=38\text{MeV}$, as given in \cite{AN}. The calculated result for
sigma term here is found to be $38.099\text{MeV and}$ listed in
\textrm{Table III}, for which $U_{0}$ is the the soliton solution of
the chirally symmetry-breaking theory (namely, the massive theory
(\ref{SK})). Being connected with the amplitude for meson-nucleon
scattering, the sigma term can be extracted from the experimental
data, with the "world average" estimate $36\pm 20\text{MeV}$
\cite{Pagels75, Reya74}, and is also listed in the table.

\section{Concluding remarks}

We show that the hybrid form of a kink-like solution and the
approximated Skyrmion profile given by the instanton method are
suited to approximate the exact profile for the symmetric Skyrmions
in the Skyrme model with pion mass term for all baryon numbers up
$B$ to $8$. This is first unified expression for the symmetric
Skyrmions for the SU(2) Skyrme model with chiral symmetry broken.
The computation of the corresponding soliton energies, and the
static properties of nucleons within the framework of collective
quantization, shows that the yielded Skyrmion profiles here are all
in a good agreement with that given by the exact numeric solution.

\section*{Acknowledgements}

D. J thanks Q. Wang and ChuengRyong Ji for discussions. This work is
supported in part by the National Natural Science Foundation of
China (No.10965005) and The Project-sponsored by SRF for ROCS, SEM.

\begin{figure*}[h]
\begin{center}
\rotatebox{0}{\resizebox *{8.5cm}{7.5cm} {\includegraphics
{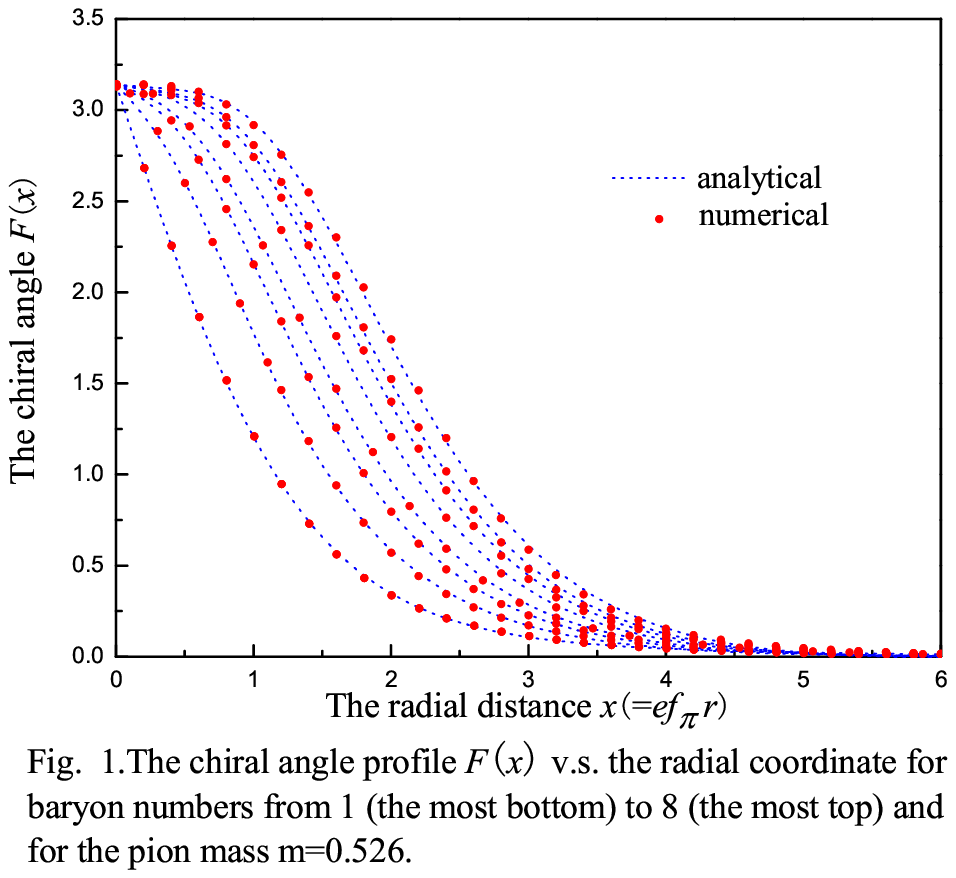}}}
\end{center}
\end{figure*}
\begin{figure*}[h]
\begin{center}
\rotatebox{0}{\resizebox *{8.5cm}{7.5cm} {\includegraphics
{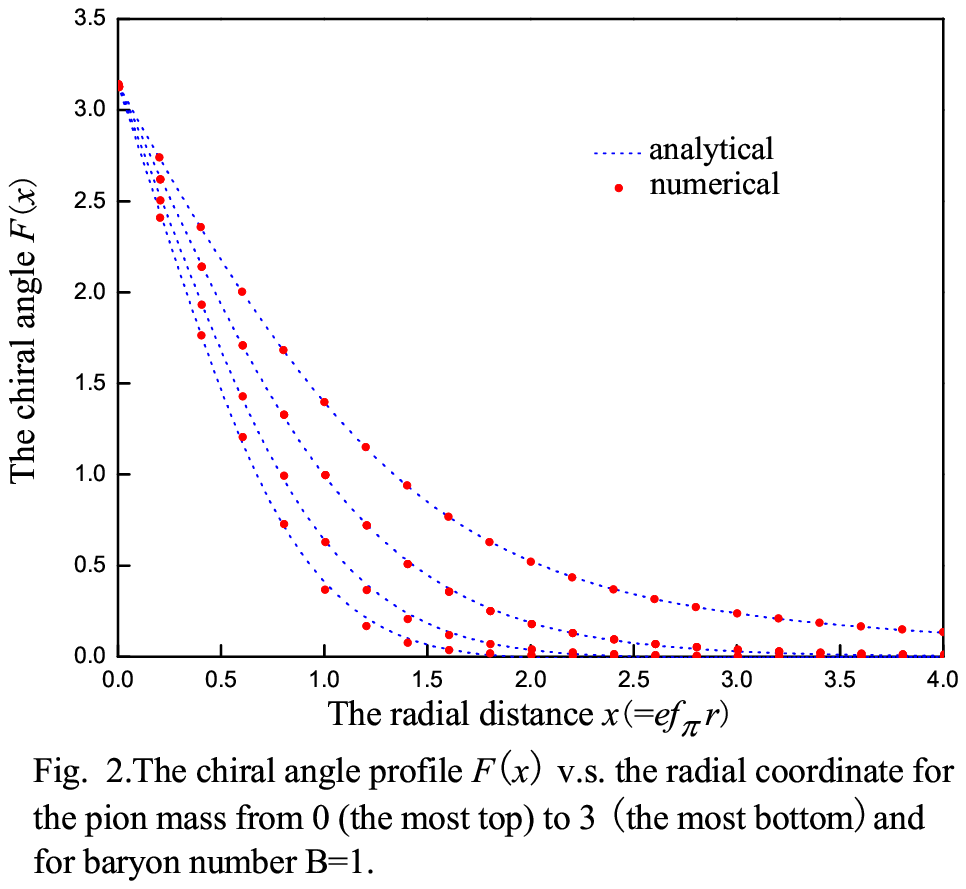}}}
\end{center}
\end{figure*}
\begin{figure*}[h]
\begin{center}
\rotatebox{0}{\resizebox *{8.5cm}{7.5cm} {\includegraphics
{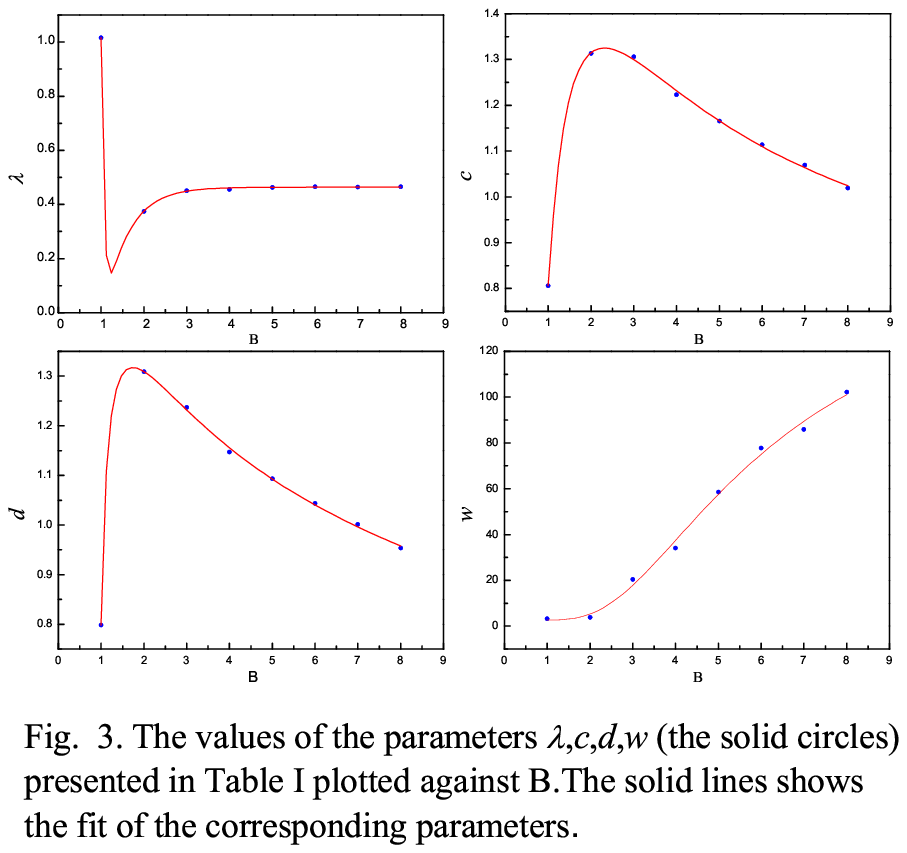}}}
\end{center}
\end{figure*}

\begin{table*}[h]
\begin{center}
\tabcolsep0.16in
\begin{tabular}{|c|c|c|c|c|c|c|c|c|}
\multicolumn{9}{c}{Table I}\\
\hline$B$&$1     $&$2     $&$3     $&$4     $&$5     $&$6     $&$7     $&$8     $\\
\hline$\mathcal{I}$&$1     $&$5.81  $&$13.58 $&$20.65 $&$35.75 $&$50.76 $&$60.87 $&$85.63 $\\
\hline$\lambda$&$1.016 $&$0.3751$&$0.4508$&$0.4558$&$0.4629$&$0.4651$&$0.4650$&$0.4658$\\
\hline$c$&$0.8058$&$1.314 $&$1.306 $&$1.223 $&$1.166 $&$1.114 $&$1.070 $&$1.019 $\\
\hline$d$&$0.7981$&$1.309 $&$1.237 $&$1.147 $&$1.093 $&$1.043 $&$1.001 $&$0.9532$\\
\hline$w$&$3.203 $&$3.760 $&$20.45 $&$34.08 $&$58.63 $&$77.72 $&$85.91 $&$102.3 $\\
\hline$\frac{E}{B}($Ana.$)$&$1.3084$&$1.2848$&$1.2603$&$1.2078$&$1.2230$&$1.2147$&$1.1817$&$1.1984$\\
\hline$\frac{E}{B}($Num.$)$&$1.3077$&$1.2833$&$1.2593$&$1.2069$&$1.2219$&$1.2132$&$1.1799$&$1.1953$\\
\hline
\end{tabular}
\end{center}
\end{table*}
\begin{table*}[h]
\begin{center}
\tabcolsep0.16in
\begin{tabular}{|c|c|c|c|}
\multicolumn{4}{c}{Table II}\\
\hline
\multicolumn{4}{|c|}{B=1,I=1}\\
\hline$m$&$1      $&$2      $&$3     $\\
\hline$\lambda$&$1.001  $&$0.9990 $&$0.9995$\\
\hline$c$      &$1.581  $&$1.813  $&$2.116 $\\
\hline$d$      &$0.8814 $&$0.5419 $&$0.4236$\\
\hline$w$      &$-0.6877$&$-1.746 $&$-2.209$\\
\hline$E($Ana.$)$ &$1.418  $&$1.647  $&$1.855 $\\
\hline$E($Num.$)$ &$1.416  $&$1.642  $&$1.847 $\\
\hline
\end{tabular}
\end{center}
\end{table*}
\begin{table*}[h]
\begin{center}
\tabcolsep0.2in
\begin{tabular}{|c|c|c|c|c|c|}
\multicolumn{6}{c}{Table III}\\
\hline Quantities &                       Ref.\cite{ANW}   &Ref.\cite{AN}   & Analytical      & Num.    & $\text{Expt.} $ \\
\hline$M_{N }($MeV$)       $ & $938.9\text{(input)} $  & $938.9\text{(input)}$ & $938.9\text{(input)}$ & $938.9\text{(input)}$ & $938.9$ \\
\hline$M_{\triangle}($MeV$)$ & $1232 \text{(input)} $  & $1232 \text{(input)}$ & $1232 \text{(input)}$ & $1232 \text{(input)}$ & $1232 $ \\
\hline$m_{\pi }($MeV$)     $ & $0    \text{(input)} $  & $138  \text{(input)}$ & $138  \text{(input)}$ & $138  \text{(input)}$ & $138  $ \\
\hline$2f_{\pi }($MeV$)                         $ & $129    $  & $108    $ & $108.20 $ & $108.27 $ & $186  $ \\
\hline$e                                        $ & $5.45   $  & $4.84   $ & $4.8424 $ & $4.8430 $ & $-    $ \\
\hline$\langle r^{2}\rangle _{I=0}^{1/2}($fm$)  $ & $0.59   $  & $0.68   $ & $0.6825 $ & $0.6821 $ & $0.72 $ \\
\hline$\langle r^{2}\rangle _{I=1}^{1/2}($fm$)  $ & $\infty $  & $1.04   $ & $1.0589 $ & $1.0408 $ & $0.88 $ \\
\hline$\langle r^{2}\rangle _{M,I=0}^{1/2}($fm$)$ & $0.92   $  & $0.95   $ & $0.9619 $ & $0.9542 $ & $0.81 $ \\
\hline$\langle r^{2}\rangle _{M,I=1}^{1/2}($fm$)$ & $\infty $  & $1.04   $ & $1.0589 $ & $1.0408 $ & $0.80 $ \\
\hline$\mu _{p}                                 $ & $1.87   $  & $1.97   $ & $1.9669 $ & $1.9664 $ & $2.79 $ \\
\hline$\mu _{n}                                 $ & $-1.31  $  & $-1.24  $ & $-1.2365$ & $-1.2369$ & $-1.91$ \\
\hline$|\mu _{p}/\mu _{n}|                      $ & $1.43   $  & $1.59   $ & $1.5906 $ & $1.5898 $ & $1.46 $ \\
\hline$\sigma ($MeV$)                           $ & $ -     $  & $38     $ & $38.099 $ & $37.726 $ & $36\pm20$\\
\hline
\end{tabular}
\end{center}
\end{table*}


\section*{References}

\begin{thebibliography}{99}
\bibitem{Skyrme} T.H.R. Skyrme, Nucl. Phys. 31(1961)556.
\bibitem{Zahed} I. Zahed\ and G. Brown, Phys. Rept. 142(1986)1.
\bibitem{Sutcliffe} P.M. Sutcliffe, Phys. Lett. B 292(1992)104.
\bibitem{Battye97} R. Battye and P.M. Sutcliffe, Phys. Rev. Lett.79%
(1997)363.
\bibitem{AtiyahManton} M.F. Atiyah and N.S. Manton, Commun. Math. Phys.
152(1993)391.
\bibitem{Atiyah89} M.F. Atiyah and N.S. Manton, Phys. Lett. B222%
(1989)438.
\bibitem{Battye} R. Battye and P.M. Sutcliffe, Phys. Rev. C73 %
(2006)055205.
\bibitem{Sextic2010} C. Adam, J.S. S\'{a}nchez-Guill\'{e}n, and
A.Wereszczynski, arXiv:1001.\\4544[hep-th]
\bibitem{ANW} G.S. Adkins, C.R. Nappi and E. Witten, Nucl. Phys. B228%
(1983)552.
\bibitem{AN} G.S. Adkins and C.R. Nappi, Nucl. Phys. B 233(1984)109.
\bibitem{JMP} J. Ananias et al., J. Math. Phys. 32, 7(1991)1949.
\bibitem{Ponciano} J.A. Ponciano et al., Phys. Rev. C64(2001)045205.
\bibitem{Yamashita} J. Yamashita and M. Hirayama, Phys. Lett. B 642%
(2006)160.
\bibitem{Faddeev} L.D. Faddeev, Lett. Math. Phys. 1(1976) 289.
\bibitem{RationM98} C.J.Houghton,N.S.Manton and P.M.Sutcliffe,Nucl.Phys.B
510(1998)\\507.
\bibitem{Manton1990} N.S.Manton,T.M.Samols,J.Phys.A23(1990)3749.
\bibitem{SKatiyah2005} M.F. Atiyah and P.M. Sutcliffe, Phys. Lett. B
605 (2005) 106--11
\bibitem{Pagels75} H. Pagels, Phys. Rept.16 (1975)219.
\bibitem{Reya74} E. Reya, Rev. Mod. Phys. 46 (1974)7159.
\bibitem{JWcpl} Duo-jie Jia, Xiao-wei Wang and Feng Liu, Submitted to Chin.
Phys. Lett. [arXiv:0912.5142].
\bibitem{RP} R.A.Battye,P.M.Sutcliffe,Rev.Math.Phys. 14(2002)29.
\end{thebibliography}
\end{document}